# New zirconium hydrides predicted by structure search method based on first principles calculations

Xueyan Zhu[a,b], De-Ye Lin[a,b], Jun Fang[a,b], Xing-Yu Gao[a,b], Ya-Fan Zhao[*a,b], Hai-Feng Song[*a,b,c]

a) Institute of Applied Physics and Computational Mathematics, Fenghao East Road 2, Beijing 100094, China

b) CAEP Software Center for High Performance Numerical Simulation, Huayuan Road 6, Beijing 100088, China

c) Laboratory of Computational Physics, Huayuan Road 6, Beijing 100088, China

* Corresponding author: zhao_yafan@iapcm.ac.cn,   song_haifeng@iapcm.ac.cn

**Abstract**

The formation of precipitated zirconium (Zr) hydrides is closely related to the hydrogen embrittlement problem for the cladding materials of pressured water reactors (PWR). In this work, we systematically investigated the crystal structures of zirconium hydride ($ZrH_x$) with different hydrogen concentrations ($x = 0\sim 2$, atomic ratio) by combining the basin hopping algorithm with first principles calculations. We conclude that the $P3m1$ $\zeta$-$ZrH_{0.5}$ is dynamically unstable, while a novel dynamically stable $P3m1$ $ZrH0.5$ structure was discovered in the structure search. The stability of bistable $P4_2/nnm$ $ZrH_{1.5}$ structures and $I4/mmm$ $ZrH_2$ structures are also revisited. We find that the $P4_2/nnm$ ($c/a > 1$) $ZrH_{1.5}$ is dynamically unstable, while the $I4/mmm$ ($c/a = 1.57$) $ZrH_2$ is dynamically stable. The $P4_2/nnm$ ($c/a < 1$) $ZrH_{1.5}$ might be a key intermediate phase for the transition of $\gamma \rightarrow \delta \rightarrow \varepsilon$ phases. Additionally, by using the thermal dynamic simulations, we find that $\delta$-$ZrH_{1.5}$ is the most stable structure at high temperature while $ZrH_2$ is the most stable hydride at low temperature. Slow cooling process will promote the formation of $\delta$-$ZrH_{1.5}$, and fast cooling process will promote the formation of $\gamma$-$ZrH$. These results may help to understand the phase transitions of zirconium hydrides.

**Key words:** Zirconium hydrides; Structure prediction; First principles calculations

# 1. Introduction

Zirconium (Zr) alloys have been used as cladding material in pressure water nuclear reactors due to their low thermal neutron absorption cross section, good mechanical properties, excellent waterside corrosion resistance, etc. [1]. However, when in long term service, complex zirconium hydrides ($ZrH_x$) will form in the cladding due to the hydrogen absorption. For decades, several zirconium hydride phases have been identified by experiment, including $\zeta$-$ZrH_{0.5}$ [2], $\gamma$-$ZrH$, $\delta$-$ZrH_{1.5}$ and $\varepsilon$-$ZrH_2$ [3].

In recent years, based on first principles calculations, thermal dynamical and mechanical properties of zirconium hydrides have been investigated, and new structures of zirconium hydrides have also been predicted [4-13]. Ackland found that for $\varepsilon$-$ZrH_2$, two structures with different $c/a$ exists and were almost degenerated in energy [14]. These bistable structures were later studied by Zhang *et al.* [15]. They suggested that the structure with $c/a > 1$ was dynamically unstable. Domain *et al.* [10] discovered two $P4_2/nnm$ $ZrH_{1.5}$ structures with different $c/a$, which are more stable than the $P\bar{n}3m$ structure. Christensen *et al.* [13] found several new structures of zirconium hydrides, including a $P\bar{n}3m$ $ZrH_{0.5}$ structure. Besson *et al.* [8] performed *ab initio* thermal dynamical simulations of fcc H-Zr system by cluster expansion method. Although much progress has been made, there lacks systematically study of the structures and phase transitions of Zr hydrides, and their stability at different temperatures still remains unknown.

Recently, crystal structure prediction methods based on first principles calculations have been developed rapidly, such as USPEX[16-17] and CALYPSO [18-19]. These methods have been successfully used to accelerate the material discoveries[18, 20]. In this work, we carried out extensive structure search calculations in order to predict the structures of $ZrH_x$, where $x$ is in the range of 0 to 2.0 (in atomic percentages). All of the structures of $ZrH_x$ observed by experiments have been reproduced, and several new zirconium hydride phases which are more stable than the known zirconium hydride phases were found. The dynamical and mechanical stability of these new structures were systematically investigated. The stability of these new structures at different temperatures was also studied.

## 2. Method and computational details

The structure search was performed with STEPMAX module of the Correlated Electron System Simulation Package (CESSP) [21-24] package, which is based on the basin hopping (BH) algorithm [25-29] for structure evolution. Geometry optimizations and energy evaluations were carried out using the projector-augmented wave (PAW) method [30] within density functional theory framework by using Vienna *ab initio* simulation package (VASP) [31-32] and CESSP. Exchange and correlation were treated in the generalized gradient approximation (GGA) as parameterized in the Perdew-Burke-Ernzerhof (PBE) [33-34] functional. The plane-wave cutoff energy was set to be 500 eV to guarantee a total energy convergence of less than 1 meV per atom. The Brillouin zone k-point sampling was performed by Monkhorst–Pack scheme [35] with the smallest allowed k-points spacing of $2\pi*0.02 Å^{-1}$. The geometry optimization is performed using the conjugate-gradient (CG) algorithm until all the forces acting on the atoms are smaller than 1.0e-4 eV/Å. The Methfessel–Paxton function [36] with a smearing width of 0.1 eV was used to perform the integration over eigenvalues. The tetrahedron method with Blöchl corrections [37] was used to obtain the accurate energies. The formation enthalpy of $Zr_{1-x}H_x$ can be calculated as

$$E^f(Zr_{1-x}H_x) = E_{Zr_{1-x}H_x} - (1-x)E_{Zr} - \frac{x}{2}E_{H_2} \quad (1)$$

where $E_{Zr_{1-x}H_x}$, $E_{Zr}$ and $E_{H_2}$ are the total energies of $Zr_{1-x}H_x$, Zr and $H_2$ respectively.

After the structure search, the dynamical stability of new structures was inverstigated by the phonon density of states (DOS). The PDOSs were calculated by using the phonopy [38] package. Furthermore, the elastic constants of these Zr hydrides were calculated. Based on these elastic constants, the mechanical stability of the predicted hybrids was examined through Born stability criteria [39-41], and their polycrystalline elastic moduli were further derived according to

Voigt-Reuss-Hill approximations [42-44].

To take into account the temperature effect, the quasi-harmonic approximation theory is used to calculate the Helmholtz free energy, and both contributions of the electron excitation free energy and the phonon are included [5]. The calculations of phonon and the thermal properties were also performed by phonopy [38] package.

### 3. Results and discussion

### 3.1 Stable and meta-stable structures of zirconium hydrides

#### 3.1.1 $ZrH_{0.5}$

For $ZrH_{0.5}$, ζ-$ZrH_{0.5}$ with *P3m1* (space group #156) symmetry has been experimentally identified and characterized by Zhao *et al.* [2]. The lattice constants were reported to be *a* = *b* = 3.3 Å and *c* = 10.29 Å. However, the stability of ζ-$ZrH_{0.5}$ remains in a dispute. Previous calculation by Zheng *et al.* [4] suggested that ζ-$ZrH_{0.5}$ is a meta-stable structure. We performed structure search calculations on $ZrH_{0.5}$ phases. In our structure search, we found several new structures which are more stable than the ζ-$ZrH_{0.5}$ phase reported by Zhao *et al.*. The formation enthalpy and lattice constants of the new structures are listed in Table 1. Since the lattice matching of zirconium hydride and α-Zr is of great importance, we also paid special attention to the structure of the Zr matrix for each zirconium hydrides. Their structures are shown in Figure 1.

Firstly, we would like to discuss the stability of the *P3m1* ζ-$ZrH_{0.5}$ structure reported by Zhao *et al*. [2]. During the structure searching, we found the reported *P3m1* structure. The calculated lattice constants (*a* = *b* = 3.26 Å, *c* = 10.85 Å) are in good agreement with previous work [2]. Zhao *et al*. suggested that one H atom occupies site (1/3, 2/3, 1/16), but in the optimized structure, the H atom actually occupies site (1/3, 2/3, -1/16). The optimized lattice constant of *c* for the *P3m1* structure predicted by us is about 5.2% larger than the experimental value of 10.29 Å.

In this study, we further checked the stability of the *P3m1* structure by phonon calculation, elastic constants calculation and more strict geometry optimization. As is shown in Figure 2a, there are several imaginary frequencies exists in the obtained phonon density of states (DOS), indicating that the *P3m1* structure is dynamically unstable. Besides, the results of elastic constants (Table 2) also showed that the P3m1 structure is mechanically unstable. Then, we optimized the *P3m1* structure with higher accuracy settings and turned the symmetry constraints off. The *P3m*1 structure transforms to a *Cm* structure ($a$ = 5.44 Å, $b$ = 3.44 Å, $c$ = 12.06 Å, $\beta$ = 118.5°). Phonon DOS simulation shows that the *Cm* structure is dynamically stable. The energy of the *Cm* structure is 30 meV per atom lower by than the original *P3m1* structure, which further proved that the *P3m1* structure is unstable.

According to Zhao *et al.* [2], ζ-ZrH$_{0.5}$ belongs to the trigonal crystal system. Through our structure searches, two new trigonal structures were found. The symmetry of one structure is $R\bar{3}m$ (space group # 166, $a = b$ = 3.28 Å, $c$ = 32.09 Å). The length of $c$ is about three times of that of the *P3m1* structure reported by Zhao *et al.* The energy of this structure is 43 meV per atom lower than the *P3m1* structure suggested by Zhao *et al.*. Besides the $R\bar{3}m$ structure, we found a new *P3m1* structure ($a = b$ = 3.26 Å, $c$ = 10.78 Å) which is different from the *P3m1* structure reported by Zhao *et al.*. The total energy of this new *P3m1* structure is 37 meV per atom lower than that of the Zhao's *P3m1* structure calculated by us. When it comes to the structure, comparing with the structure reported by Zhao *et al.*, the lattice constant of $c$ is 0.06 Å shorter, and it is also closer to the experimental value, and the positions of H atoms in the unit cell are different as it was mentioned above.

The most stable structure of ZrH$_{0.5}$ predicted in our structure search is a cubic crystal with $Pn\bar{3}m$ symmetry (space group #224, $a$ = 4.648 Å), which is in agreement with the results by Christensen *et al.* [13]. The Zr matrix of the $Pn\bar{3}m$ structure is of $Fm\bar{3}m$ symmetry with $a$ = 4.648 Å. Based on our calculations, the total energy of the $Pn\bar{3}m$ structure is 78 meV/atom lower than the *P3m1* structure reported by Zhao *et al.*

However, the $Pn\bar{3}m$ ZrH$_{0.5}$ structure has not been observed by experiment. Thus, we further investigated the stability of $Pn\bar{3}m$ ZrH$_{0.5}$ at higher temperature by the quasi-harmonic approximation calculations of Helmholtz free energy. As shown in Figure 3a, with the increase of temperature, the relative stability of the $Pn\bar{3}m$ structure decreases, while the *Cmmm* and *C2/m* (S1 and S2) structures of ZrH$_{0.5}$ become more stable. The *Cmmm* and the *C2/m* (S2) structures turn to be more stable than the $Pn\bar{3}m$ structure when the temperature is higher than 900K, while *C2/m* (S1) becomes as stable as the $Pn\bar{3}m$ structure at 1100K. The new *P3m1* is also getting more stable as the temperature increases, but it is still less stable than the $Pn\bar{3}m$ structure even at 2000K. The existence of ζ-ZrH$_{0.5}$ is possibly due to the lattice matching between α-Zr which stabilizes the Zr/ζ-ZrH$_{0.5}$ interface, the high transformation barrier to other more stable structures, or the existence of alloying elements in the commercial zirconium alloys used in experimental observations.

Through the analysis of the above results, we conclude that the ζ-ZrH$_{0.5}$ structure reported by Zhao *et al.* is actually unstable. We found a new *P3m1* structure which is more stable. However, the most stable ZrH$_{0.5}$ structure predicted by us is a $Pn\bar{3}m$ structure, but it might be less stable at high temperature.

### 3.1.2. ZrH$_{1.0}$

For ZrH, the γ-ZrH with tetragonal cell has been experimentally observed, but its stability is still in a dispute. Though γ-ZrH was previously reported to be of *P4$_2$/n* symmetry with $a = b = 4.586$ Å, $c = 4.948$ Å (space group # 86) [45] with four Zr atoms and four H atoms in a unit cell, its actual symmetry is *P4$_2$/mmc* (space group # 131) [46] with a Zr$_2$H$_2$ unit cell. A *Cccm* structure was also reported by Kolesnikov *et al.* in 1994 [47], with lattice constants being $a = 4.549$ Å, $b = 4.618$ Å and $c = 4.965$ Å. This structure is usually regarded as a distorted structure of the *P4$_2$/mmc* phase [3].

In this work, the most stable ZrH structure predicted at 0 K is also P4$_2$/mmc. The optimized lattice constants are $a = 3.237$ Å and $c = 5.004$ Å, which is in a good agreement with previous DFT result [46]. Though a *Cccm* structure with its lattice constants being very close to the values reported by Kolesnikov [47] was found during the structure search, further geometry optimization simulation with tighter convergence criteria showed that its actual structure is still of *P4$_2$/mmc* symmetry.

The second and third most stable structures of ZrH predicted by us are of *Ccce* and *P*222 symmetry. The total energies of these structures are 24.1 and 25.4 meV per atom higher than the *P4$_2$/mmc* structure, respectively. Their lattice constants are listed in Table 1.

Weck *et al.* also reported two *P*222 structures [46] which are different from the *P*222 structure found in our calculations. As a comparison, we optimized the Weck's structures and found that their symmetries are actually *P4$_2$22* and $P\bar{4}2m$. Their lattice constants are $a = b = 4.888$ Å, $c = 4.410$ Å for *P4$_2$22* structure and $a = b = 4.778$ Å, c = 4.641 Å for $P\bar{4}2m$ structure, respectively. The energies of the two structures are 17.7 and 20.5 meV higher than the *P*222 structure by our calculations.

As is shown in Figure 2b, phonon density of states of *P4$_2$/mmc*, *Ccce* and *P*222 structures have no imaginary frequencies. This indicates that these structures are dynamically stable. Furthermore, as shown in Table 2, the calculated elastic constants also show that the *P4$_2$/mmc* structure is mechanically stable.

The finite temperature effect on the structural stability of ZrH was investigated and result is displayed in Figure 3b. As temperature increases, the free energies of the *Ccce* and *P*222 structure are getting closer to the *P4$_2$/mmc* structure, but the *P4$_2$/mmc* structure remains to be the most stable structure even up to 2000K.

As a summary, the *P4$_2$/mmc* structure is the most stable ZrH structure at low temperature as well as high temperature based on our calculations.

### 3.1.3 ZrH$_{1.5}$

For ZrH$_{1.5}$, the experimentally observed structure is a face-centered cubic (*fcc*) phase with lattice constants of 4.768 Å. By using a unit cell with 10 atoms (Zr$_4$H$_6$), we find a cubic structure of $Pn\bar{3}m$ symmetry (space group #224, $a = 4.775$ Å), which agrees with experimental measurement and previous DFT calculations by Zhu *et al.*[48].

However, the *fcc* structure is not the most stable one according to our results of structure search. Based on our calculations, the most stable structure of ZrH$_{1.5}$ is a *P4$_2$/nnm* tetragonal structure with lattice constants of $a = b = 5.020$ Å and $c = 4.261$

Å ($c/a < 1$). The Zr matrix of this $P4_2/nnm$ structure is of $I4/mmm$ symmetry with lattice constants of $a = 3.550$ Å, $c = 4.261$ Å, $c/a = 1.20$. It is worth mentioning that this structure is one of the two structures of $ZrH_{1.5}$ reported by Domain *et al.* [10].

We also build the second $P4_2/nnm$ structure according to the study of Domain *et al.* by hand. The optimized lattice constants of this structure are $a = b = 4.604$ Å and $c = 5.110$ Å ($c/a > 1$), and its structural energy is 22.6 meV per atom higher than the $P4_2/nnm$ ($c/a < 1$) structure. Nevertheless, the $P4_2/nnm$ ($c/a>1$) structure is still 8.2 meV/atom more stable than the $P n\bar{3}m$ structure. Furthermore, as shown in figure 2(c), the calculated phonon density of state has several minor imaginary frequencies, which indicates that the $P4_2/nnm$ ($c/a > 1$) structure suggested by Domain *et al.* is dynamically unstable. A small displacement was applied to the second $P4_2/nnm$ structure, and after geometry optimization we got the most stable $P4_2/nnm$ structure as we predicted.

These two $P4_2/nnm$ structures are not the only structures more stable than the *fcc* $ZrH_{1.5}$. We also find another three structures with symmetries of $P\bar{1}$, $P2/c$ and *Ibam*, which are more stable than the $P n\bar{3}m$ structure. Lattice parameters of these structures and some other structures such as the $P4_2/mcm$ and $P\bar{4}m2$ which have also been reported by Weck *et al.*[46] and Zheng *et al.*[4] are also summarized in Table 1.

The above discussions indicate that there are many structures that are more stable than the experimentally observed $P n\bar{3}m$ at 0 K. To explain why the experimentally observed structure is $P n\bar{3}m$, we considered the temperature effect on these structures. As shown in Figure 3c, at 0 K, the $P n\bar{3}m$ phase is less stable than $P4_2/nnm$, $P\bar{1}$, $P2/c$ and *Ibam* phases; as the temperature increases, the $P n\bar{3}m$ phase is getting more stable; at 600 K, the $P n\bar{3}m$ phase becomes more stable than $P\bar{1}$, $P2/c$ and *Ibam* phases; when the temperature is higher than

1000 K, the $Pn\bar{3}m$ phase becomes more stable than the *P4$_2$/nnm* structure.

In summary, the *P4$_2$/nnm* ( *c/a* < 1 ) is the most stable ZrH$_{1.5}$ structure at 0 K, while the *P4$_2$/nnm (c/a>1)* structure is dynamically unstable. At high temperature (~1000K), the $Pn\bar{3}m$ structure becomes to be the most stable ZrH$_{1.5}$ structure.

### 3.1.4 ZrH$_2$

For ZrH$_2$, ε-ZrH$_2$ with *I4/mmm* symmetry (space group #139) and fcc-ZrH$_2$ with $Fm\bar{3}m$ symmetry (space group #225) have been extensively studied. Both structures were also found in our calculations. The *I4/mmm* structure is predicted to be most stable. Its lattice constants are *a* = *b* = 3.538 Å, *c* = 4.406 Å, *c/a* = 1.25, which agrees well with the previous calculation by Chattaraj *et al.* [49]. We also find another I4/*mmm* structure with lattice constants of *a* = 3.278 Å and *c* = 5.150 Å, *c/a* =1.570. It can be regarded as a distorted structure of *I4/mmm* (*c/a* = 1.25). Its formation energy is only 2.6 meV per atom higher than ε-ZrH$_2$.

The bistable structures of *I4/mmm* ZrH$_2$ have also been discussed by Ackland *et al.* [14] and Zhang *et al.* [15]. Zhang *et al.* reported that ZrH$_2$ phase has two distinct *fct* structures with *c/a* = 0.885 and *c/a* = 1.111, which corresponds to the *I4/mmm* (*c/a* = 1.25) and I4/*mmm* (*c/a* = 1.57) predicted by our calculations, respectively. They concluded that the *fct*-0.885 structure is more stable than the *fct*-1.111 one. Though pervious calculation by Zhang *et al.* suggested that the *fct*-1.111 *I4/mmm* structure is dynamically unstable, our calculation shows no imaginary vibration frequencies in the simulated phonon DOS for *I4/mmm* (*c/a* = 1.570). The contradictory conclusions obtained by us and Zhang *et al.* is possibly due to the more strict convergence criteria adopted in our calculations.

The $Fm\bar{3}m$ structure of ZrH$_2$ found by us was with the lattice constant *a* = 4.807 Å, which is in good agreement with previous calculations by Zhang *et al.* [15]. The energy of this $Fm\bar{3}m$ structure is 9.2 meV per atom higher than the stable *I4/mmm* structure. However, its phonon DOS has a small imaginary vibrational frequency (Figure 2(d)), indicating that the $Fm\bar{3}m$ structure is dynamically unstable. This is also consistent with the previous work of Zhang *et al.*[15]. By applying random small displacements to the $Fm\bar{3}m$ structure and followed by geometry optimization, we got the ZrH$_2$ of *I4/mmm* (*c/a* = 1.25) structure. As shown in Table 2, mechanical property calculations showed that the $Fm\bar{3}m$ structure is also mechanically unstable. Considering that the $Fm\bar{3}m$ structure is very similar to $Pn\bar{3}m$ ZrH$_{1.5}$ structure, it is possible that the $Fm\bar{3}m$ structure is a transition state in the transition path between $Pn\bar{3}m$ ZrH$_{1.5}$ and *I4/mmm* ZrH$_2$.

A $R\bar{3}m$ structure with lattice constants of *a* = 3.461 Å and *c* = 8.026 Å was found to be more stable than the $Fm\bar{3}m$ structure. The $R\bar{3}m$ structure is 6.6 meV per atom higher in energy than ε-ZrH$_2$, but it is still 2.6 meV per atom more stable than the $Fm\bar{3}m$ structure. It is possible that the $R\bar{3}m$ structure is also a metastable structure which might be observed experimentally under some conditions.

Furthermore, we investigated the finite temperature effect on the different ZrH$_2$ structures. As shown in Figure 2 and Table 3, the $Fm\bar{3}m$ structure is both dynamically and mechanically unstable, but its Helmholtz free energy is lower than the *I4/mmm* structure as the temperature is higher than 900 K. However, this structure could be transformed into *I4/mmm* structure under lower temperature, because the $Fm\bar{3}m$ structure is dynamically unstable.

In summary, two *I4/mmm* structure different *c/a* (1.25 and 1.57) are the two most stable ZrH$_2$ structures. Both of them are dynamically stable.

**3.2 Phase transitions of the zirconium hydrides**

Based on the experimental observations, as H atoms accumulated, the structures of the zirconium hydrides changes from ζ-ZrH$_{0.5}$, γ-ZrH, δ-ZrH$_{1.5}$ to ε-ZrH$_2$. Their symmetry varies from *P3m*1 (trigonal), *P4$_2$/mmc* (tetragonal), $Pn\bar{3}m$ (cubic) and finally to *I4/mmm* (tetragonal). By removing the hydrogen atoms from these structures, we analyzed the structural changes of their Zr matrixes. The Zr matrix of ζ-ZrH$_{0.5}$ is still of *P3m*1 symmetry, but it is very close to the α-Zr with *P6$_3$/mmc* symmetry. For γ-ZrH, the Zr matrix is of the *I4/mmm* symmetry with *c/a*~1.55. For δ-ZrH$_{1.5}$, the Zr matrix is of the $Fm\bar{3}m$ symmetry. For ε-ZrH$_2$, the Zr matrix is changed again to be of the *I4/mmm* symmetry, but the *c/a*~1.24. Therefore, the structure of Zr matrix changes drastically with the increasing of hydrogen concentration.

However, considering the new structures we found in the structure searching, the structure transitions of γ, δ and ε phases could be explained more clearly.

As is shown in Figure 4, the hydrogen absorption of γ-ZrH leads to the formation of *P4$_2$/nnm* (*c/a* <1) ZrH$_{1.5}$, and their corresponding Zr matrix changes from *I4/mmm*(*c/a*~1.55) to I4/mmm(*c/a*~1.20). The *P4$_2$/nnm* (*c/a* > 1) ZrH$_{1.5}$ is a possible transition state of the hydrogenation reaction. The *P4$_2$/nnm* (*c/a* < 1) ZrH$_{1.5}$ structure could be transformed to ZrH$_{1.5}$ via hydrogen diffusion. The further hydrogenation of *P4$_2$/nnm*(*c/a* < 1) ZrH$_{1.5}$ leads to the formation of *I4/mmm*(*c/a* = 1.25*) ZrH$_2$. The symmetries of corresponding Zr matrix of *P4$_2$/nnm*(*c/a* < 1) and *I4/mmm*(*c/a* = 1.25) both are *I4/mmm*(*c/a*~1.20). The *P4$_2$/nnm*(*c/a* < 1) of ZrH$_{1.5}$ may be a key reaction intermediate in the phase transition of Zr hydrides.

### 3.3 Stability of Zr hydrides at different temperature

In our global minima searching process, the formation enthalpies at different temperature are evaluated at 0 K. Whether these structures remain stable at higher temperature is still an open question. Here we investigated the effect of temperature based on the QHA method.

We also investigated the formation free energy of the following bulk reactions via hydrogen diffusion at different temperatures:

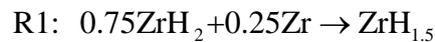

R1:  $0.75 ZrH_2 + 0.25 Zr \rightarrow ZrH_{1.5}$

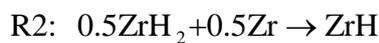

R2:  $0.5 ZrH_2 + 0.5 Zr \rightarrow ZrH$

R3: $0.25ZrH_2 + 0.75Zr \rightarrow ZrH_{0.5}$

R4: $ZrH_{1.5} + 0.5Zr \rightarrow 1.5ZrH$

R5: $ZrH_{1.5} + 2Zr \rightarrow 3ZrH_{0.5}$

R6: $2ZrH_{1.5} \rightarrow ZrH_2 + ZrH$

R7: $3ZrH_{1.5} \rightarrow 2ZrH_2 + ZrH_{0.5}$

R8: $ZrH + Zr \rightarrow 2ZrH_{0.5}$

R9: $2ZrH \rightarrow ZrH_{1.5} + ZrH_{0.5}$

R10: $3ZrH \rightarrow ZrH_2 + 2ZrH_{0.5}$

For $ZrH_2$, $ZrH_{1.5}$ and ZrH, the free energy of experimentally observed ε-$ZrH_2$ (*I4/mmm*), δ-$ZrH_{1.5}$ ($Pn\bar{3}m$) and γ-ZrH(*P4$_2$/mmc*) are used, while for $ZrH_{0.5}$, the free energy of the new *P3m*1 structure found by our structure search is used. Results of the above reactions are illustrated in Figure 5. At low temperature (less than 300 K), reactions of R1, R2, R3, R5, R8, R9 and R10 are endothermic, thus these reactions cannot happen spontaneously. However, reactions of R4 and R6 are exothermic and can happen spontaneously. It can be concluded the stability order of these four hydrides are $ZrH_2$, ZrH, $ZrH_{1.5}$, and $ZrH_{0.5}$.

As the temperature increases, the free energies of some reactions change. At 1100K, reactions of R4 and R6 turn from exothermic at low temperature to endothermic, which means that $ZrH_{1.5}$ is more stable than ZrH. Meanwhile, R1 turns from endothermic to exothermic, indicating that $ZrH_{1.5}$ is also more stable than $ZrH_2$. The stability order of these four hydrides becomes $ZrH_{1.5}$, $ZrH_2$, ZrH and $ZrH_{0.5}$.

From the above analysis, we can see that as temperature increases, δ-$ZrH_{1.5}$ becomes the most stable structure. By fast cooling process, Zr hydride will keep the high temperature structure unchanged, which means that δ-$ZrH_{1.5}$ should be the structure observed after fast cooling. At low temperature, reaction R4 becomes exothermic, which means that slow cooling will promote the formation of γ-ZrH. Our conclusions agree with the recent experiment by Wang *et al.*[50]. The transition of $ZrH_{1.5}$ to $ZrH_2$ and Zr (reverse reaction of R1) is also exothermic. This reaction is also observed by recent experiment by Maimaitiyili *et al*[51]. As suggested by Wang *et al.*, the transition from δ to γ or ε is diffusion controlled. Since the diffusion of H in δ-$ZrH_{1.5}$ is much slower than in α-Zr [52], γ-ZrH is more likely to be formed than ε-$ZrH_2$.

For the formation of $ZrH_{0.5}$, the only possible reactions are R7 at low temperature. In this reaction, $ZrH_{1.5}$ decomposes to be $ZrH_2$ and $ZrH_{0.5}$. In the experiment of Zhao *et al.*, the specimens were annealed at 550 °C for 1 h and then slowly cooled to room temperature. Our simulation also agrees with this experiment. Though reactions of R3, R5, R8, R9 and R10 are all exothermic at low temperature, they might be kinetically prohibited by the high hydrogen diffusion barriers in the Zr hydrides.

**4. Conclusions**

Our conclusions could be summarized as follows.

1. We have systematically performed structure searching for $ZrH_x$ with different hydrogen concentrations, where $x$=0.5, 1.0, 1.5, and 2.0. All the previously reported structures of zirconium hydrides have been reproduced. We found a new $P3m1$ $ZrH_{0.5}$ structure, which is much more stable than the $P3m1$ ζ-$ZrH_{0.5}$ reported by Zhao *et al.*.

2. Phonon DOS were calculated to investigate the stabilities of these structures. The results show that the $P3m1$ ζ-$ZrH_{0.5}$ reported by Zhao *et al.*, $P4_2/nnm$ ($c/a > 1$) $ZrH_{1.5}$ and the $Fm\bar{3}m$ $ZrH_2$ are dynamically unstable, but the $I4/mmm$ ($c/a \sim 1.5$) $ZrH_2$ is dynamically stable.

3. The stability of Zr hybrids at different temperatures was investigated by extensive calculations of the entropy contributions to the Helmholtz free energy. Results show that though the $Pn\bar{3}m$ structure is stable at low temperature, its stability decrease at high temperature. The $Pn\bar{3}m$ $ZrH_{1.5}$ is more stable than the $P4_2/nnm$ structures at 1000 K.

4. By the calculations of the elastic constants, the mechanic stability of the

predicted hybrids was determined according to the Born criteria. It was found that the $Pn\bar{3}m$ ZrH$_{0.5}$, $P4_2/nnm$ ($c/a < 1$) ZrH$_{1.5}$ and $I4/mmm$ ($c/a \sim 1.25$) ZrH$_2$ are mechanically stable, while $P3m1$ ζ-ZrH$_{0.5}$, $P4_2/nnm$ ($c/a > 1$) ZrH$_{1.5}$, $I4/mmm$ ($c/a \sim 1.55$) ZrH$_2$ and $Fm\bar{3}m$ ZrH$_2$ are mechanically unstable.

5. The $P4_2/nnm$ ($c/a < 1$) structure might be a key intermediate for the phase transition from γ to δ and ε phase.

6. By investigating the reaction free energy of several conversion reactions between the hydrides, we concluded that ZrH$_2$ is the most stable Zr hydride at low temperature and $Pn\bar{3}m$ δ-ZrH$_{1.5}$ is the most stable Zr hydride at high temperature. Fast cooling of high temperature Zr hydrides leads to formation of δ-ZrH$_{1.5}$ and slow cooling leads to the formation of γ-ZrH.

**Acknowledgement**

The authors acknowledge funding support from National Key Research and Development Program of China (under Grants No. 2016YFB0201203）and National High Technology Research and Development Program of China under Grant No. 2015AA01A304. We also thank Dr. Yan-Chao Wang at Jilin University for valuable discussions.

Table 1 Structure symmetry and formation enthalpy ($E_F$) of different zirconium hydrides at 0 K

| Compounds | Symmetry | $E^f$(eV) | Lattice Constants(Å; °) | Zr matrix structure | Note |
|---|---|---|---|---|---|
| $ZrH_{0.5}$ | $Pn\bar{3}m$ | -0.256 | $a = 4.647$ | $Fm\bar{3}m$ ($a = 4.647$) | |
| | Cmmm | -0.244 | $a = 6.496, b = 9.553, c = 3.227$ | I4/mmm ($a = 3.238, c = 4.777$) | |
| | C2/m | -0.244 | $a = 5.603, b = 3.309, c = 22.086; \beta = 102.79$ | $P\bar{3}m1$ ($a = 3.27, c = 21.54$) | C2/m S1 |
| | C2/m | -0.243 | $a = 10.858, b = 3.219, c = 12.988; \beta = 153.96$ | $Fm\bar{3}m$ ($a = 4.637$) | C2/m S2 |
| | $R\bar{3}m$ | -0.221 | $a = 3.279, c = 32.089$ | $R\bar{3}m$ ($a = 3.279, c = 32.089$) | |
| | P3m1 | -0.215 | $a = 3.262, c = 10.788$ | P3m1 ($a = 3.262, c = 10.788$) | New P3m1 |
| | P3m1 | -0.178 | $a = 3.262, c = 10.855$ | $P\bar{6}m2$ ($a = 3.262, c = 10.855$) | P3m1 by Zhao et al. |
| ZrH | $P4_2/mmc$ | -0.430 | $a = 3.233, c = 5.016$ | I4/mmm ($a = 3.233, c = 5.016$) | |
| | Ccce | -0.406 | $a = 6.978, b = 6.939, c = 8.622$ | I4/mmm ($a = 3.479, c = 4.311$) | |
| | P222 | -0.405 | $a = 4.632, c = 9.815$ | $Fm\bar{3}m$ ($a = 4.723$) | |
| $ZrH_{1.5}$ | $P4_2/nnm$ | -0.511 | $a = 5.020, c = 4.261$ | I4/mmm ($a = 3.550, c = 4.261$) | fct, $c/a < 1$ |
| | $P4_2/nnm$ | -0.506 | $a = 4.607, c = 5.104$ | I4/mmm ($a = 3.258, c = 5.104$) | fct, $c/a > 1$ |
| | $P\bar{1}$ | -0.506 | $a = 3.260, b = 5.629, c = 6.001; \alpha=85.03, \beta=94.08, \gamma=80.02$ | I4/mmm ($a = 3.526, c = 4.329$) | |
| | P2/c | -0.505 | $a = 3.287, b = 5.033, c = 6.577; \beta = 81.31$ | I4/mmm ($a = 3.543, c = 4.284$) | |
| | Ibam | -0.505 | $a = 4.750, b = 9.622, c = 4.768$ | $Fm\bar{3}m$ ($a = 4.776$) | |
| | $Pn\bar{3}m$ | -0.503 | $a = 4.774$ | $Fm\bar{3}m$ ($a = 4.774$) | |
| | *Fmmm* | -0.501 | $a = 9.779$, b = 6.669, c = 6.675 | $Fm\bar{3}m$ ($a = 4.775$) | |
| | $P4_2/mcm$ | -0.499 | $a = 4.768$, c = 4.794 | $Fm\bar{3}m$ ($a = 4.777$) | |
| | $P\bar{4}m2$ | -0.496 | $a = 3.348$, c = 4.868 | $Fm\bar{3}m$ ($a = 4.779$) | |
| $ZrH_2$ | *I4/mmm* | -0.573 | $a = 3.538$, c = 4.406 | *I4/mmm* ($a = 3.538$, $c = 4.406$) | fct, $c/a=1.25$ |
| | *I4/mmm* | -0.570 | $a = 3.278$, c = 5.150 | *I4/mmm* ($a = 3.278$, $c = 5.150$) | fct, $c/a=1.57$ |
| | $R\bar{3}m$ | -0.566 | $a = 3.461$, c = 8.026 | $Fm\bar{3}m$ ($a = 4.810$) | |
| | $Fm\bar{3}m$ | -0.564 | $a = 4.807$ | $Fm\bar{3}m$ ($a = 4.807$) | |

Table 2 Elastic constants and polycrystalline elastic moduli (GPa) of zirconium hydride, and their comparisons with DFT results from the literatures.

| | Symmetry | $C_{11}$ | $C_{12}$ | $C_{13}$ | $C_{33}$ | $C_{44}$ | $C_{14}$ | $C_{66}$ | $B$ | $G$ | $E$ | Mechanical stability |
|---|---|---|---|---|---|---|---|---|---|---|---|---|
| $ZrH_{0.5}$ | $Pn\bar{3}m$ | 118.586 | 96.116 | 96.116 | 118.586 | 59.843 | – | 59.843 | 103.606 | 31.158 | 84.957 | √ |
| | P3m1 | 159.774 | 83.103 | 60.204 | 190.655 | 34.267 | 2.842 | 38.336 | 101.896 | 40.614 | 107.553 | √ |
| | P3m1 [3] ($\zeta$-$ZrH_{0.5}$) | -20.21 | 248.47 | 67.11 | 185.81 | -144.8 | -48.04 | -134.34 | 100.99 | -287.86 | -17302 | × |
| ZrH | $P4_2/mmc$ | 116.266 | 112.431 | 97.337 | 183.002 | 49.383 | – | 61.114 | 112.973 | 23.711 | 66.483 | √ |
| | $P4_2/mmc$ [3] | 119.89 | 115.55 | 97.39 | 176.58 | 48.35 | – | 57.99 | 114.40 | 23.55 | 66.13 | |
| | $P4_2/mmc$ [4] ($\gamma$-ZrH) | 122 | 116 | 98 | 183 | 47.6 | – | 61.1 | 116 | 25.3 | 69.5 | |
| $ZrH_{1.5}$ | $P4_2/nnm$ | 159.615 | 139.044 | 102.706 | 114.902 | 58.573 | – | 29.707 | 118.578 | 28.953 | 80.321 | √ |
| | $Pn\bar{3}m$ | 89.046 | 142.72 | – | – | 33.413 | – | 33.413 | 124.829 | 168.467 | 348.586 | × |
| | $Pn\bar{3}m$ [3] ($\delta$-$ZrH_{1.5}$) | 94.52 | 138.30 | – | – | 31.07 | – | – | 123.71 | 486.62 | 631.64 | |
| $ZrH_2$ | I4/mmm | 174.325 | 147.351 | 104.425 | 153.148 | 31.604 | – | 65.816 | 132.632 | 31.508 | 87.588 | √ |
| | I4/mmm [3] | 163.11 | 146.84 | 107.28 | 146.01 | 29.22 | – | 63.01 | 130.72 | 26.12 | 73.48 | |
| | I4/mmm [4] | 166 | 149 | 109 | 149 | 26.5 | – | 55.8 | 133 | 24.9 | 70.1 | |

| | | | | | | | | | | |
|---|---|---|---|---|---|---|---|---|---|---|
| I4/mmm [36] (ε-ZrH$_2$) | 172.3 | 149.9 | 104.4 | 145.5 | 31.8 | – | 63.8 | – | – | – |
| I4/mmm | 126.047 | 149.602 | 115.124 | 196.101 | 27.97 | – | 44.746 | 133.532 | 77.832 | 195.51 | × |
| I4/mmm [35] | 125.7 | 145.5 | 115.0 | 190.6 | 30.9 | – | 42.0 | – | – | – |
| Fm$\bar{3}$m | 71.72 | 165.111 | – | – | -36.994 | – | – | 133.98 | -40.611 | -135.525 | × |
| Fm$\bar{3}$m [3] | 78.06 | 160.35 | – | – | -39.42 | – | – | 132.92 | -40.10 | -133.75 |
| Fm$\bar{3}$m [35] (fcc-ZrH$_2$) | 82.6 | 159.7 | – | – | -19.5 | – | – | – | – | – |

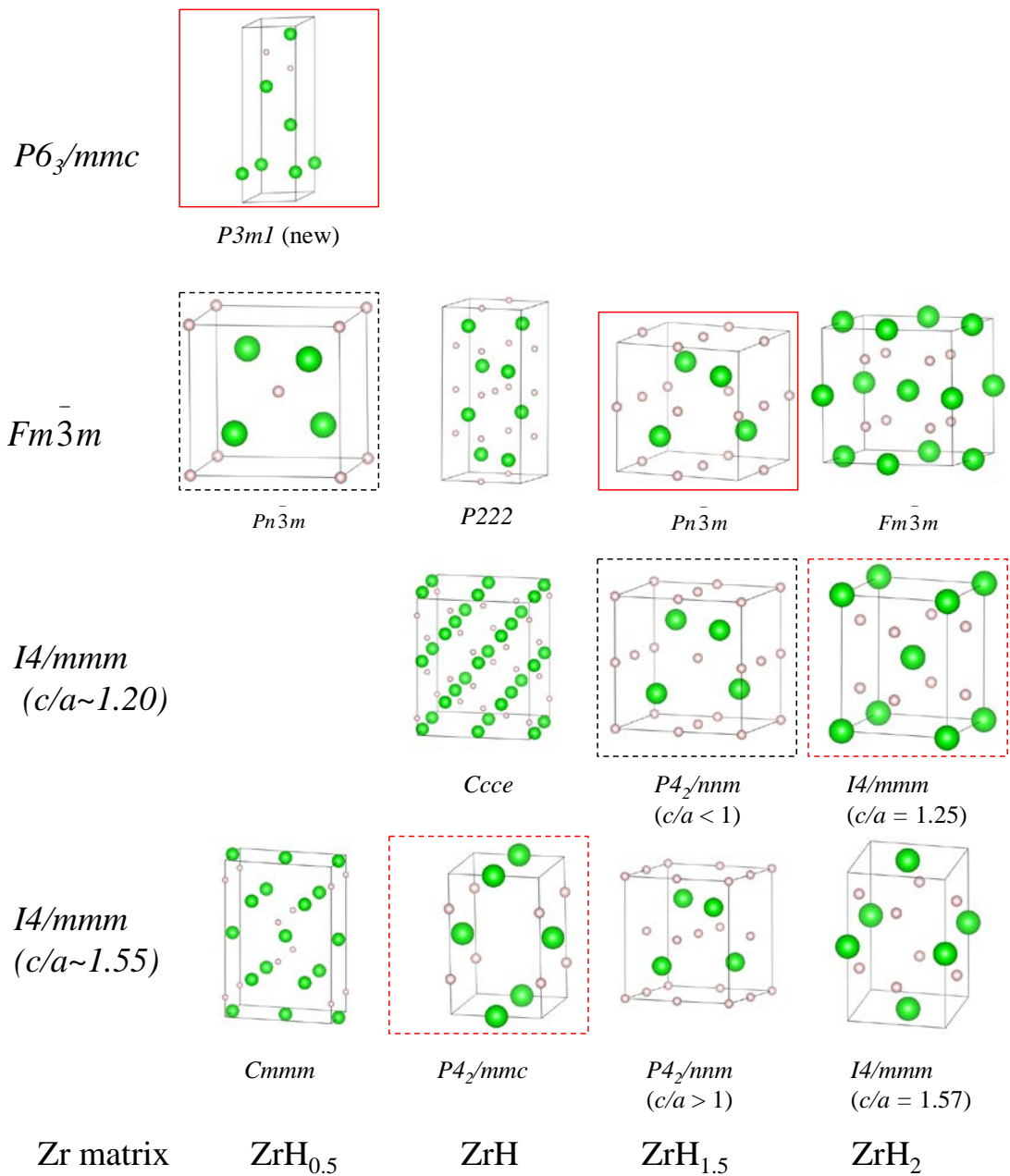

**Figure 1** Structure of zirconium hydrides. The most stable structures given by *ab initio* simulations are noted with dotted squares and the experimental observed structures are noted by red squares.

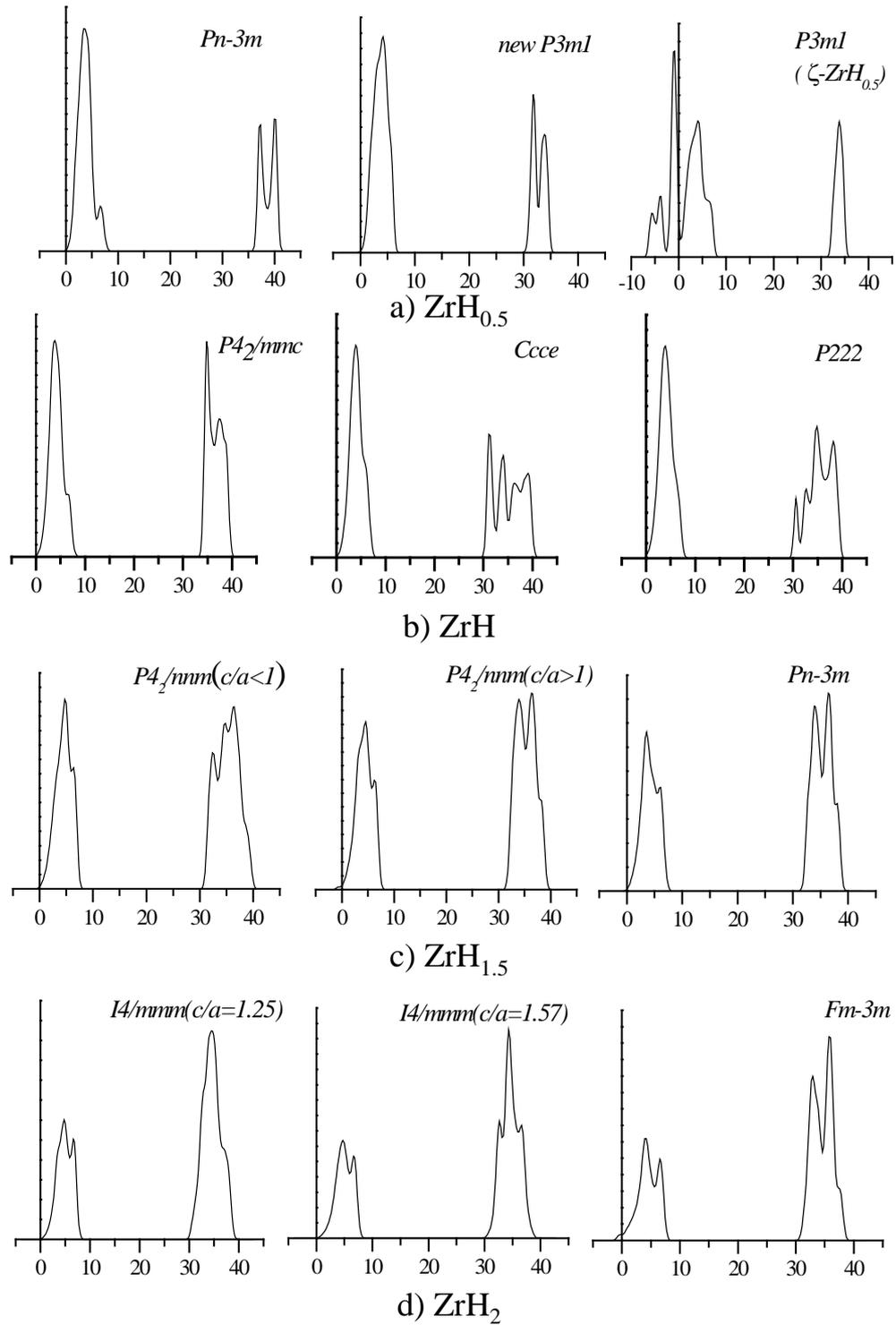

**Figure 2** Phonon density of states of $ZrH_{0.5}$, ZrH, $ZrH_{1.5}$ and $ZrH_2$ structures

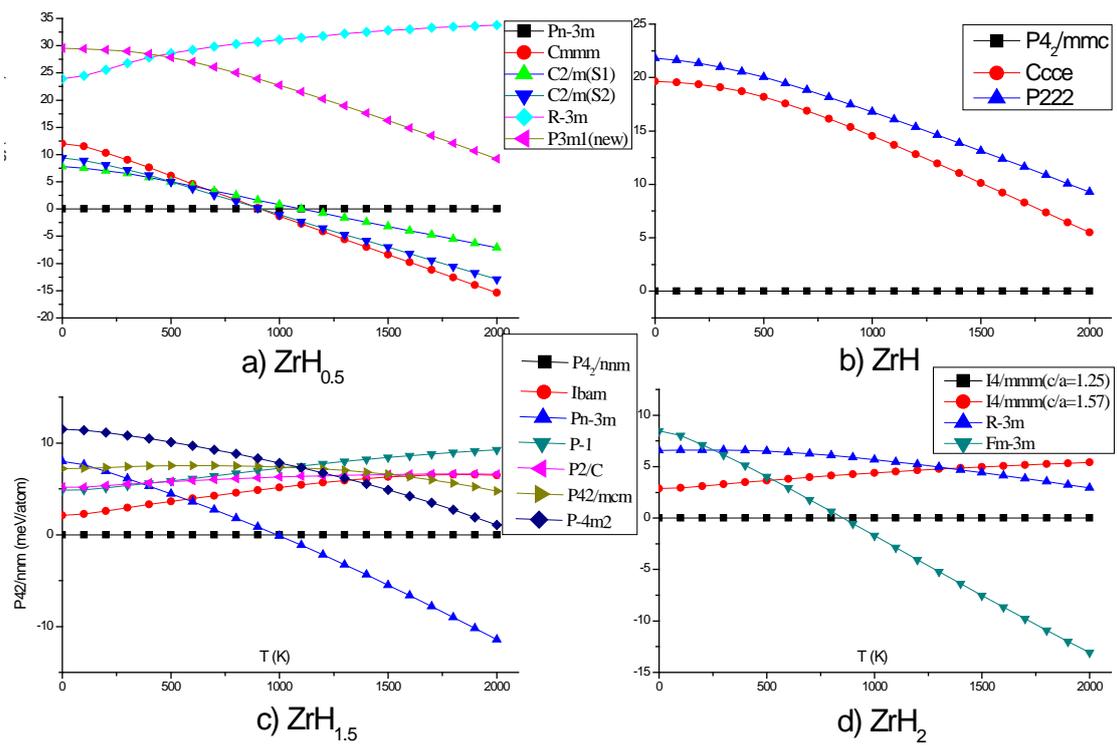

**Figure 3** Helmholtz free energy of different structures

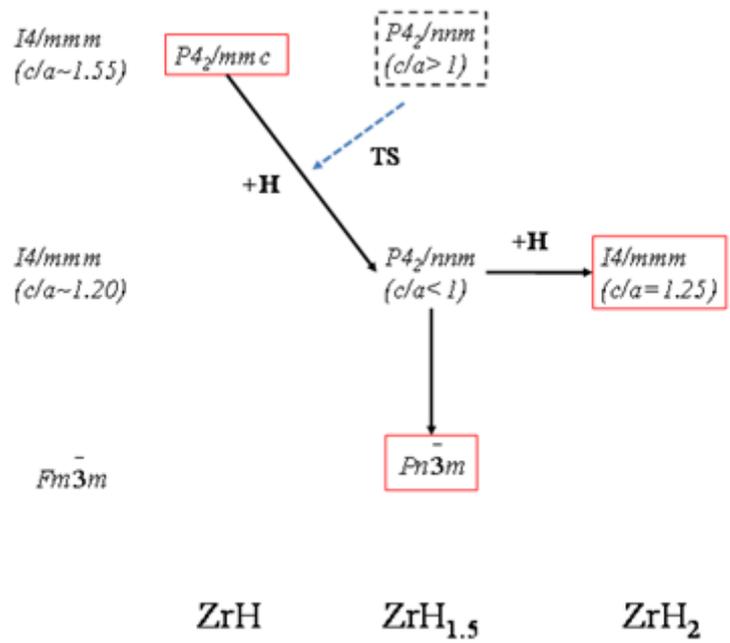

**Figure 4** Phase transition of γ-ZrH, δ-ZrH$_{1.5}$ and ε-ZrH$_2$

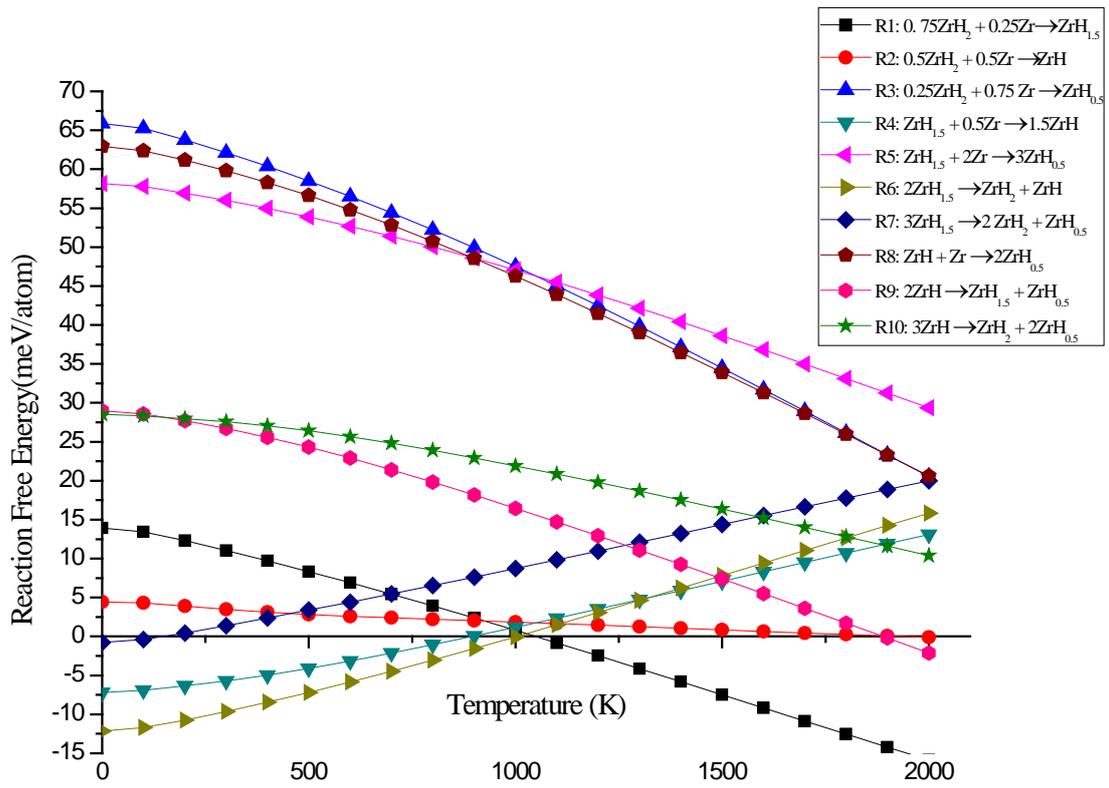

**Figure 5**  Reaction free energy for zirconium hydrides at different temperature